\documentclass[10pt,conference]{IEEEtran}
\IEEEoverridecommandlockouts
\makeatother
\usepackage[x11names]{xcolor}
\usepackage{cite}
\usepackage{amsmath,amssymb,amsfonts}
\usepackage{algorithmic}
\usepackage{graphicx}
\usepackage{textcomp}
\usepackage{bbm}
\usepackage{caption}
\usepackage{booktabs}
\usepackage{multirow}
\usepackage{array}
\usepackage{arydshln}
\usepackage{enumitem}
\usepackage{pifont}
\usepackage[colorlinks,linkcolor=green!50!black,urlcolor=blue!70!black,citecolor=red!60!black]{hyperref}
\usepackage{marvosym}
\usepackage{todonotes}
\usepackage{lipsum}
\usepackage{tcolorbox}
\usepackage{threeparttable}
\def\BibTeX{{\rm B\kern-.05em{\sc i\kern-.025em b}\kern-.08em
    T\kern-.1667em\lower.7ex\hbox{E}\kern-.125emX}}
\begin{document}

\title{The Seeds of the FUTURE Sprout from History: \\Fuzzing for Unveiling Vulnerabilities in Prospective Deep-Learning Libraries}

\author{
    Zhiyuan Li${}^{\dagger\ddagger}$, Jingzheng Wu${}^{\dagger\S\P}$\textsuperscript{\Letter}, Xiang Ling${}^{\dagger\S\P}$\textsuperscript{\Letter}, Tianyue Luo${}^{\dagger}$, Zhiqing Rui${}^{\dagger\ddagger}$, Yanjun Wu${}^{\dagger\S\P}$ \\[1ex]
    \begin{tabular}{c}
        ${}^{\dagger}$Institute of Software Chinese Academy of Sciences \\ 
        ${}^{\ddagger}$University of Chinese Academy of Sciences \\
        ${}^{\S}$Key Laboratory of System Software (Chinese Academy of Sciences)\\
        ${}^{\P}$State Key Laboratory of Computer Science, Institute of Software Chinese Academy of Sciences
    \end{tabular} \\
    \thanks{\textsuperscript{\Letter}Jingzheng Wu and Xiang Ling are the corresponding authors.}
}

\maketitle

\begin{abstract}
The widespread application of large language models (LLMs) underscores the importance of deep learning (DL) technologies that rely on foundational DL libraries such as PyTorch and TensorFlow. Despite their robust features, these libraries face challenges with scalability and adaptation to rapid advancements in the LLM community. In response, tech giants like Apple and Huawei are developing their own DL libraries to enhance performance, increase scalability, and safeguard intellectual property. Ensuring the security of these libraries is crucial, with fuzzing being a vital solution. However, existing fuzzing frameworks struggle with target flexibility, effectively testing bug-prone API sequences, and leveraging the limited available information in new libraries. To address these limitations, we propose FUTURE, the first universal fuzzing framework tailored for newly introduced and prospective DL libraries. FUTURE leverages historical bug information from existing libraries and fine-tunes LLMs for specialized code generation. This strategy helps identify bugs in new libraries and uses insights from these libraries to enhance security in existing ones, creating a cycle from history to future and back. To evaluate FUTURE's effectiveness, we conduct comprehensive evaluations on three newly introduced DL libraries. Evaluation results demonstrate that FUTURE significantly outperforms existing fuzzers in bug detection, success rate of bug reproduction, validity rate of code generation, and API coverage. Notably, FUTURE has detected 148 bugs across 452 targeted APIs, including 142 previously unknown bugs. Among these, 10 have been assigned CVE IDs. Additionally, FUTURE detects 7 bugs in PyTorch, demonstrating its ability to enhance security in existing libraries in reverse. 
\end{abstract}


\section{Introduction}
Artificial intelligence (AI) continues to lead technological innovation, with large language models (LLMs) rapidly gaining widespread application~\cite{aillm}. Within the pipeline of AI, deep learning (DL) has made significant strides~\cite{branch}, prompting a growing demand for efficient and versatile programming frameworks. This demand has led to the development of DL libraries (\emph{e.g.}, PyTorch~\cite{pytorch} and TensorFlow~\cite{tensorflow}). Despite their comprehensive capabilities, these libraries face limitations in scalability, flexibility, and proprietary control, which are critical for tech giants like Apple and Huawei. To address these limitations and meet the complex requirements of LLMs, these companies invest heavily in developing their own DL libraries.

Emerging DL libraries, including Apple MLX~\cite{mlx2023}, Huawei MindSpore~\cite{mindspore}, and OneFlow~\cite{oneflow}, aim to improve performance through enhanced computational efficiency and advanced features. However, their complexity may introduce bugs that pose critical risks in sectors like healthcare~\cite{healthcare}, finance~\cite{finance}, and autonomous driving~\cite{auto}. Identifying and addressing bugs in these libraries is essential for the safety of downstream DL systems. Despite high level of attention and adoption these libraries have received in their early development stages, there is a noticeable lack of mature and comprehensive testing methodologies. This gap underscores the need to ensure their security and reliability.

Fuzzing has proven highly effective in detecting software bugs by automatically generating test cases that expose unexpected behaviors~\cite{fuzzing}. Existing fuzzing methods for DL libraries are mainly categorized into API-level fuzzing~\cite{docter,freefuzz,nablafuzz,tensorscope} and model-level fuzzing~\cite{cradle,audee,muffin}. API-level fuzzing focuses on individual API functions to uncover bugs triggered by anomalous API inputs but may miss complex bugs due to its inability to construct intricate API sequences~\cite{apisequence}. Model-level fuzzing tests entire models against inputs that exploit architectural or weight bugs. Although it addresses many limitations of API-level fuzzing, model-level fuzzing covers a limited range of APIs~\cite{limitedapi}. Recent works (\emph{e.g.}, TitanFuzz~\cite{titanfuzz} and FuzzGPT~\cite{fuzzgpt}) leverage LLMs to address some limitations of API-level fuzzing methods. While API-level, model-level, and recent LLM-based methods have demonstrated excellent performance in detecting bugs within DL libraries, applying these methods to newly introduced and prospective DL libraries poses several challenges:

\noindent \textbf{C1: Lack of Target Flexibility.} Existing fuzzing methods, primarily designed for PyTorch and TensorFlow, lack the flexibility needed to adapt to new libraries. Applying these methods to new libraries requires extensive resource collection and code modifications, which are both time-consuming and labor-intensive. Despite this fundamental limitation, other challenges persist within the existing methods, as outlined below.

\noindent \textbf{C2: Lack of Bug-prone API Sequences.} Previous API-level methods~\cite{docter,freefuzz,nablafuzz,tensorscope} struggle to effectively test API sequences. TitanFuzz~\cite{titanfuzz} leverages LLMs to address this limitation. However, in essence, TitanFuzz's generation and mutation still rely on the inherent capabilities of LLMs, resulting in the random generation of multi-API code snippets. This approach fails to test bug-prone API sequences, leading to an excessively large search space and limited efficiency in finding bugs.

\noindent \textbf{C3: Limited Available Information.} Existing methods utilize documentation~\cite{docter}, open-source code snippets~\cite{freefuzz}, and historical bugs~\cite{fuzzgpt} to guide the fuzzing process. However, newly introduced and prospective libraries typically only provide documentation, sometimes including code examples, in their early stages. The availability of open-source code snippets and historical bugs is extremely limited. Therefore, documentation becomes the primary resource, but existing methods that rely on it, such as Docter~\cite{docter}, are limited to testing individual APIs, as mentioned in \textbf{C2}. 

\noindent \textbf{C4: Lag in Knowledge of LLMs.} Despite the rapid updates in LLMs, it typically takes several months for the latest models to incorporate knowledge about newly introduced libraries. As a result, LLM-based fuzzers (\emph{e.g.}, TitanFuzz~\cite{titanfuzz}, FuzzGPT~\cite{fuzzgpt} and Fuzz4All~\cite{fuzz4all}) that rely on pre-trained models are significantly constrained in their effectiveness.

To overcome these challenges, we propose FUTURE, the first universal fuzzing framework tailored for both newly introduced and prospective DL libraries. Existing libraries contain a wealth of historical bug information, including complex and bug-prone API sequences that closely mimic real-world usage scenarios. The basic idea of FUTURE is to leverage the historical bug information from existing libraries (referred to as source libraries) to identify bugs in newly introduced and prospective libraries (referred to as target libraries). Firstly, we design a label-specific GitHub spider to crawl issues related to bugs in source libraries. We extract codes that reproduce these issues as historical bug codes (see Section~\ref{sec:Historical Bug Collection}). To bridge the gap between source and target libraries, we utilize a universal prompt template to leverage the limited information available in the API documentation and code examples of the target libraries. Using prompts crafted from this template, we invoke LLMs to generate code pairs that illustrate how source and target libraries implement the same API functions. FUTURE then mutates these code pairs to construct datasets for fine-tuning. With the fine-tuned LLMs, we generate seed codes by converting historical bug codes from source to target libraries and generating codes that invoke the API of the target libraries (see Section~\ref{sec:Seed Code Generation}). Finally, we conduct differential testing on three newly introduced libraries to demonstrate the effectiveness of FUTURE. We identify bugs, extract bug-prone API inputs, and test these inputs on the source libraries to unveil undetected bugs (see Section~\ref{sec:Test Oracle}). In summary, FUTURE makes the following contributions:

\begin{itemize}[leftmargin=0.5cm]
\item \textbf{Universal Fuzzing Framework:} We propose FUTURE, the first universal fuzzing framework for both newly introduced and prospective DL libraries. This framework significantly reduces the effort required to adapt fuzzing techniques to any new DL library, making it a forward-thinking tool. Additionally, FUTURE is the first fuzzing method that targets Apple MLX.
\item \textbf{From History to FUTURE and Back:} We establish a code conversion mapping between existing and prospective libraries by fine-tuning LLMs. This mapping allows us to convert historical bug codes from existing libraries into seed codes for prospective libraries, pioneering a fuzzing method from historical insights to future anticipations. Furthermore, by leveraging bugs found in prospective libraries, FUTURE enhances security in existing libraries, completing a cycle from future back to history.
\item \textbf{Bug Detection:} FUTURE demonstrates remarkable efficacy by detecting 148 bugs across 452 targeted APIs in Apple MLX, Huawei MindSpore, and OneFlow, including 142 previously unknown bugs. Among these bugs, 10 have been assigned CVE IDs. In addition, FUTURE has identified 7 bugs in PyTorch. We release the details of detected bugs and FUTURE implementation at \href{https://github.com/Redmept1on/FUTURE}{https://github.com/Redmept1on/FUTURE}.
% Besides bugs, FUTURE detects 20 enhancement issues and 12 documentation problems. 
\end{itemize}

\section{Background and Related Work}
\subsection{Deep learning and DL libraries}

In the burgeoning field of AI, DL represents a paradigmatic shift, employing models that emulate the intricate neural structures of the human brain~\cite{deeplearning}. This advanced branch of machine learning (ML) relies on neural networks to identify subtle patterns in large datasets, supporting decision-making and eliminating the need for manual feature extraction~\cite{patterns}. 

DL libraries serve as pivotal components within DL systems, providing tools and interfaces for designing, implementing, and efficiently training neural models. They simplify complex mathematical operations and hardware interactions, making deep learning more accessible to developers and researchers~\cite{interface,deepsec,lingsurvey,jiang,lingwolf}. Early libraries like Theano~\cite{theano} and Caffe~\cite{caffe} established the foundation, while PyTorch~\cite{pytorch} and TensorFlow~\cite{tensorflow} revolutionized the landscape with features like automatic differentiation~\cite{Automaticdifferentiation} and adaptive computational graphs~\cite{adaptivegraph}, accelerating research and model refinement.

Recent developments include libraries like Apple MLX~\cite{mlx2023}, Huawei MindSpore~\cite{mindspore} and OneFlow~\cite{oneflow}, which optimize performance and scalability for DL tasks. These libraries enhance computational efficiency in large-scale and distributed environments, addressing the needs of an era marked by exponentially growing data volumes, model complexities, and the development of LLMs. The widespread deployment of these libraries provides essential infrastructure for developing DL models across various sectors including healthcare~\cite{healthcare}, autonomous vehicles~\cite{selfdriving} and finance~\cite{finance}. Consequently, it is vital to maintain rigorous oversight to ensure these libraries meet high standards of quality and security.

\begin{figure*}[ht]
  \centering
  \includegraphics[width=\linewidth]{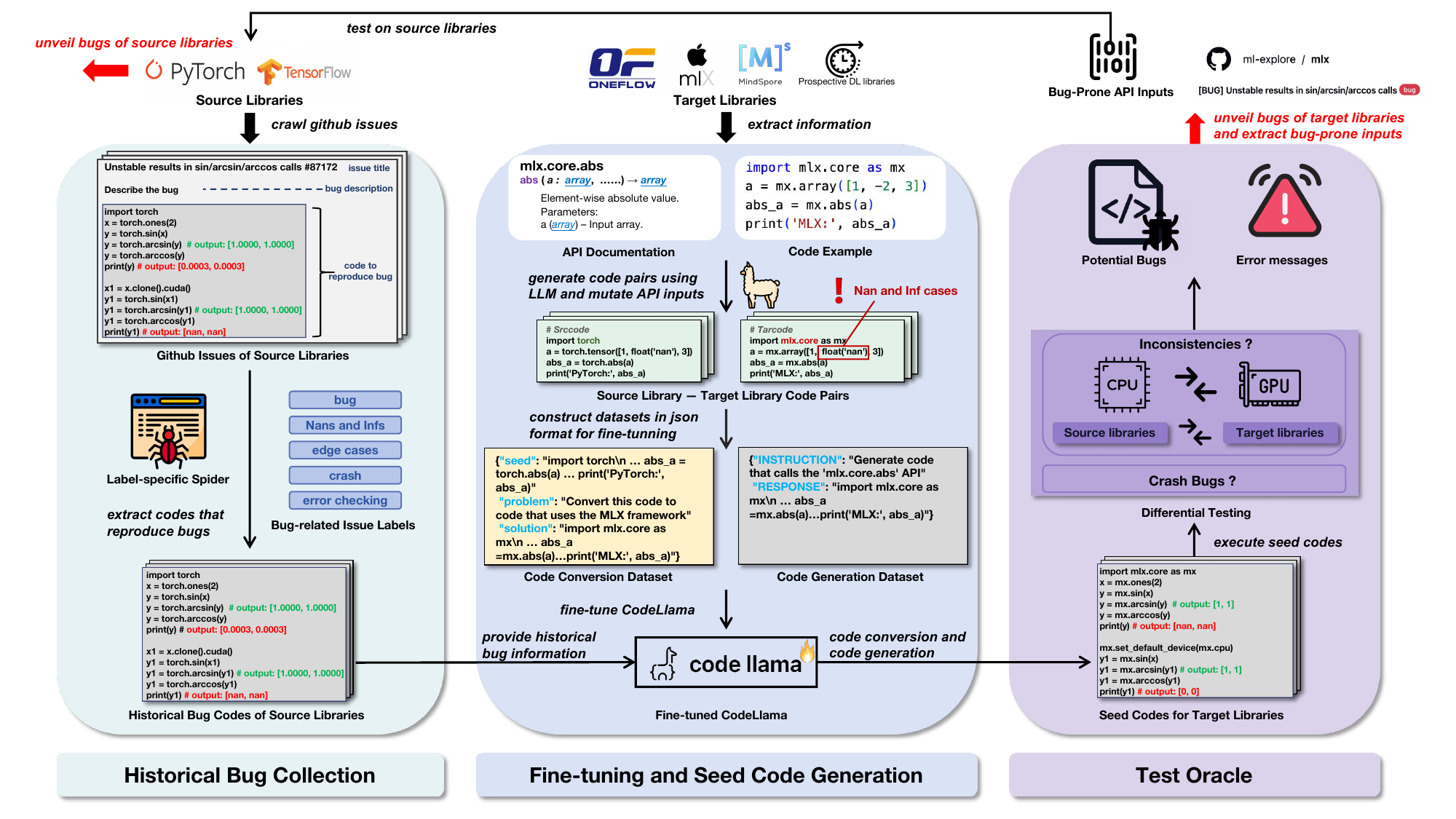}
  \caption{\textbf{Overview of FUTURE.} FUTURE leverages historical bug information from source libraries and available API information from target libraries to realize code pairs generation, dataset construction, LLM fine-tuning, and seed code generation. Utilizing these seed codes, FUTURE unveils bugs in target libraries through test oracle. Insights gained from these bugs are used to enhance the security of the source libraries, completing a cycle from history to future and back.}
  \label{img:overview}
\end{figure*}

\subsection{Fuzzing with Large Language Models}
The emergence of LLMs has transformed various domains, providing unprecedented capabilities in text generation~\cite{textgen}, code generation~\cite{codegen}, and more~\cite{llmsurvey,llmapplication,gpt,yuxiao}. These capabilities are revolutionizing fields such as content creation~\cite{contentcreation}, conversational AI~\cite{conversationai}, and software testing~\cite{gptsoftwaretest}.

Pre-trained LLMs (\emph{e.g.}, GPT-3~\cite{gpt3} and Llama~\cite{llama}) serve as the foundational models, offering broad capabilities across diverse linguistic tasks due to their extensive training on text data. General-purpose fine-tuned LLMs (\emph{e.g.}, Codex~\cite{codex} and CodeLlama~\cite{codellama}) extend these models by focusing on specific capabilities like coding generation, making them versatile tools in various applications~\cite{generalfine}. However, they sometimes lack the nuanced understanding required for highly specialized tasks.

Task-specialized fine-tuned models such as CodeLlama-Python and CodeLlama-Instruct~\cite{codellamapython} address this limitation by tailoring general-purpose fine-tuned models to excel in particular domains. These models integrate domain-specific knowledge during further fine-tuning, enabling them to solve problems that pre-trained and general-purpose fine-tuned models struggle with, such as understanding complex domain jargon or predicting highly specialized outcomes in fields like medical diagnostics or financial forecasting.

Integrating LLMs into the fuzzing process allows fuzzers to explore systems in ways traditional methods cannot, investigate potential failure modes, and ensure robustness against a wider range of input scenarios~\cite{failuremode}. LLMs are extensively employed in fuzzing frameworks for compilers, protocols, and various other domains~\cite{compilerfuzz,protocolfuzz,driverfuzz}. Several frameworks have utilized LLMs to enhance DL library fuzzing. TitanFuzz~\cite{titanfuzz} is the first approach that directly leverages a general-purpose fine-tuned LLMs (Codex and Incoder) to generate and mutate DL programs for fuzzing. However, due to limitations in model capability and knowledge lag, TitanFuzz cannot effectively test complex, bug-prone API sequences or provide immediate testing for newly introduced libraries. FuzzGPT~\cite{fuzzgpt} demonstrates that LLMs can be prompted or fine-tuned to resemble historical bug-triggering programs. However, FuzzGPT is limited by its reliance on historical bug-triggering programs from the libraries being tested, which is impractical for newly introduced and prospective libraries. Fuzz4All~\cite{fuzz4all} effectively utilizes LLMs for input generation and mutation across various software systems under test (SUTs), however, when applied to DL libraries, it encounters the same challenges as TitanFuzz.

\section{Approach}\label{sec:Approach}
\subsection{Overview}
As shown in Fig.~\ref{img:overview}, FUTURE comprises three main phases: historical bug collection, fine-tuning and seed code generation, and test oracle.

\begin{figure*}[ht]
  \centering
  \includegraphics[width=\linewidth]{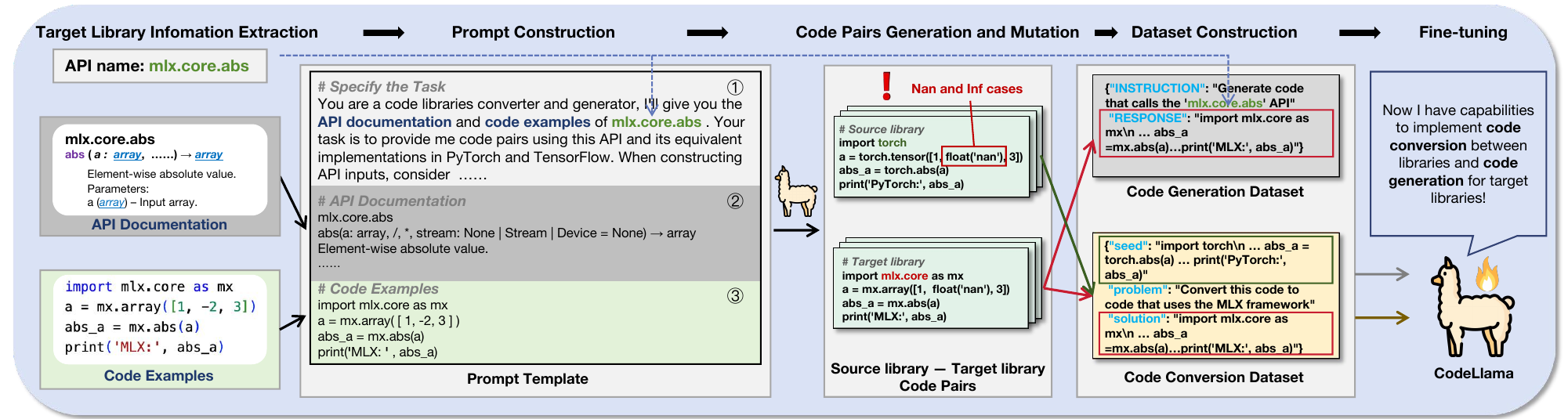}
  \caption{\textbf{Datasets Construction and Fine-tuning.} In the prompt template, we emphasize the importance of constructing API inputs that consider NaNs and Infs, edge cases, and scenarios likely to trigger API error checking and crashes. Due to space constraints, these details are omitted in the figure.}
  \label{img:finetune}
\end{figure*}

In the historical bug collection phase (Section~\ref{sec:Historical Bug Collection}), we focus on existing libraries. To gather historical bug information, we design a label-specific spider to crawl bug-related GitHub issues, retrieving code snippets that trigger bugs. These snippets serve as historical bug codes. 

In the fine-tuning and seed code generation phase (Section~\ref{sec:Seed Code Generation}), we concentrate on newly introduced and prospective DL libraries. We first collect API documentation and code examples (Section~\ref{sec:Target Library Information Extraction}). Then, we construct a prompt template to leverage these resources (Section~\ref{sec:GPT Prompt Construction}). By invoking LLMs with prompts, we generate and mutate code pairs that consist of an API call in a target library and its corresponding implementation in a source library (Section~\ref{sec:Code Pairs Generation and Mutation}). The obtained code pairs are then converted into datasets to fine-tune LLMs (Section~\ref{sec:Dataset Construction}). With the fine-tuned LLMs, we convert the historical bug codes into codes using the target libraries. These converted codes, along with codes randomly generated by fine-tuned LLMs for the target libraries, form the seed codes (Section~\ref{sec:Fine-tuning and Seed Code Generation}). 

In the test oracle phase (Section~\ref{sec:Test Oracle}), we execute the seed codes to perform differential testing on the target libraries, identifying potential bugs through abnormal behaviors such as crashes and inconsistencies. Utilizing the bugs detected by FUTURE in the target libraries, we extract API inputs that are likely to trigger bugs and attempt to detect unaddressed bugs in the source libraries. 

\subsection{Historical Bug Collection}\label{sec:Historical Bug Collection}

Several existing DL libraries (\emph{e.g.}, PyTorch~\cite{pytorch} and TensorFlow~\cite{tensorflow}) have significant user engagement, resulting in thousands of bug reports. Users typically interact with project developers via GitHub issues, reporting potential bugs. These reports include basic descriptions, system environment information, and code snippets to reproduce the issues. Developers address these issues by committing fixes for verified bugs and labeling them accordingly.

Inspired by this process, to collect historical bug codes, we design a fully automated, label-specific spider. This spider parses issue pages, categorizes issues by labels, retrieves issue contents, extracts code snippets and perform preprocessing.

Specifically, our spider traverses each page of issues, retrieving the title and ID of each entry. It subsequently accesses the issue pages to extract code snippets tagged with keywords such as ``Standalone code to reproduce the issue", ``Usage example" or ``Code example". These extracted snippets are then processed to enhance usability by importing necessary dependencies and removing non-code elements. These snippets are then saved, with directory and file names generated based on the issue's title and ID. To ensure valid filenames, special characters are replaced, and a numerical suffix is appended in cases of duplicate names. The saved snippets serve as FUTURE's historical bug codes, enabling systematic organization and storage for further analysis and examination.

\subsection{Fine-tuning and Seed Code Generation}\label{sec:Seed Code Generation}

\subsubsection{Target Library Information Extraction}\label{sec:Target Library Information Extraction}
Firstly, FUTURE extracts API documentation and code examples from the official documentation of target libraries. For each API, the API name \begin{math}Name\end{math}, API documentation \begin{math}Doc\end{math}, and code examples \begin{math}Code\end{math} are saved in a triplet format \begin{math}\mathsf{APIinfo}=\left\{Name, Doc, Code \right\}\end{math}. For APIs lacking documentation and referencing other APIs, FUTURE extracts information from the referenced APIs to ensure comprehensive coverage. 

\subsubsection{Prompt Construction}\label{sec:GPT Prompt Construction} 
After retrieving \begin{math}\mathsf{APIinfos}\end{math}, we devise a universal prompt template to construct prompts. The constructed prompts consist of three parts:

\noindent \textbf{Specify the Task.} This part defines the task to obtain code pairs that consist of an API call in a target library and its corresponding implementation in a source library, more details of this part are shown in Fig.~\ref{img:finetune} (\textcircled{1}) .

\noindent \textbf{API Documentation.} For APIs with extensive documentation, we streamline the prompts by removing redundant information to stay within the \begin{math}max\_tokens\end{math} limit of LLMs.

\noindent \textbf{Code Examples.} For code examples that present multiple usage methods for a single API, we extract different methods. By deconstructing the code examples, we derive new prompts \begin{math}P^{'}\end{math} from the original prompts \begin{math}P\end{math}. To be more specific, we first use regular expressions~\cite{regular} to determine whether the \begin{math}Name\end{math} appears multiple times in the code example. If the \begin{math}Name\end{math} appears only once, we directly use the code example to construct the prompt (i.e., \begin{math}p^{'}\end{math} = \begin{math}p\end{math}). If the \begin{math}Name\end{math} appears multiple times, we use abstract syntax trees~\cite{ast} to decompose and reassemble the code example, splitting \begin{math}k\end{math} usage methods into multiple separate code examples, thereby decomposing \begin{math}p\end{math} into \begin{math}p^{'} = \{p_{1}, p_{2}, \ldots, p_{k}\}\end{math}, where \begin{math}p \in P\end{math} and \begin{math}p^{'} \subseteq P^{'}\end{math}. Fig.~\ref{img:finetune} (\textcircled{2},\textcircled{3}) illustrates the incorporation of API documentation and code examples into our prompt template.

\subsubsection{Generation and Mutation of Code Pairs}\label{sec:Code Pairs Generation and Mutation} 

Subsequently, we query LLMs with new prompts \begin{math}P^{'}\end{math} to obtain code pairs \begin{math}CP = \left\{CP_1, CP_2, \ldots, CP_n \right\}\end{math}, where \begin{math}n\end{math} represents the number of prompts eventually constructed. Each code pair \begin{math}CP_i=(S_i, T_i), i \in (1, 2, 3, \ldots, n)\end{math} in \begin{math}CP\end{math} consists of source library code \begin{math}S\end{math} and corresponding target library code snippets \begin{math}T\end{math}. To manage costs, we limit the number of code pairs generated per API. We then randomly select variables in the API input data and mutate them to random numbers. Assuming that each code pair \begin{math}CP_i\end{math} undergoes \begin{math}m\end{math} mutations, \begin{math}n*m\end{math} mutated code pairs are acquired, which can be represented as \begin{math}CP^{'} = \{CP_{11}, CP_{12}, \ldots,\end{math} \begin{math} CP_{1m} , \ldots, CP_{nm}\}\end{math}. By deconstructing code examples, generating code pairs and mutating them, we acquire multiple code pairs for a single API, resulting in more robust data for fine-tuning.

\begin{figure}[t]
  \centering
  \includegraphics[width=\linewidth]{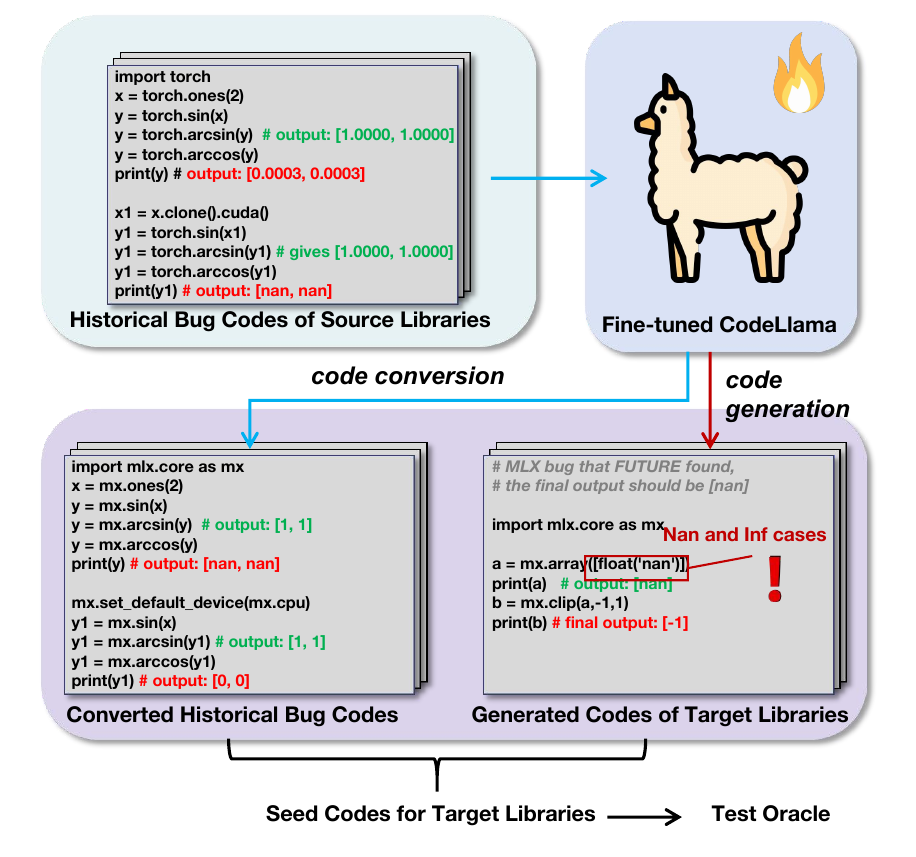}
  \caption{\textbf{Seed Code Generation.} With the task-specialized fine-tuned LLMs, we perform code conversion and code generation to obtain the seed codes for test oracle.}
  \label{img:conv_and_gen}
\end{figure}

\subsubsection{Dataset Construction}\label{sec:Dataset Construction}

After obtaining \begin{math}CP^{'}\end{math}, we convert them into datasets for fine-tuning, resulting in two datasets:

\noindent \textbf{Code Generation Dataset:} Since LLMs are not pre-trained with knowledge of prospective libraries, our goal is to equip them to generate code for the target libraries. Therefore, we construct the code generation dataset using \begin{math}T\end{math} as the ``RESPONSE", providing standard answers for code generation, as illustrated in Fig.~\ref{img:finetune}. 

\noindent \textbf{Code Conversion Dataset:} To equip LLMs with the capability to convert code snippets from the source libraries to target libraries, we construct a triplet format dataset using \begin{math}CP^{'}\end{math}. This dataset designates \begin{math}S\end{math} as ``seed" and \begin{math}T\end{math} as ``solution", with the ``problem" explicitly stated as ``Convert this code to code that uses the target library (MLX/MindSpore/OneFlow)". 

\subsubsection{Fine-tuning and Seed Code Generation}\label{sec:Fine-tuning and Seed Code Generation}
We design a universal fine-tuning template to enable users to adapt FUTURE to their own requirements with minimal effort. There are three popular fine-tuning techniques for pre-trained models~\cite{pre-trained}: Fine Tuning~\cite{fine}, Parameter-efficient Fine-Tuning (PEFT)~\cite{peft}, and Prompt Tuning~\cite{prompttuning}. We employ the Low-Rank Adaptation (LoRA)~\cite{lora} from PEFT in the template.

Consider \(W\in\mathbb{R}^{d \times k}\) as the weight matrix of LLMs. Instead of updating \(W\) directly, LoRA introduces two low-rank matrices \(A\in\mathbb{R}^{d \times r} \) and \( B \in \mathbb{R}^{r \times k} \), where \( r\ll \min(d, k)\). The weight update is then parameterized as $\Delta W = A B$. During fine-tuning, the effective weight matrix \(W'\) is given by: \[W' = W + \Delta W = W + A B\]
The low-rank matrices \( A \) and \( B \) are updated during the fine-tuning process, while the original weight matrix \( W \) remains fixed. The update rule for \(A\) and \(B\) typically follows the gradients of the loss function \( L \) with respect to these matrices. Let \(\theta = \{ A, B \} \) denote the parameters of the low-rank matrices. The gradient descent update step for \( \theta \) is given by:
\[\theta \leftarrow \theta - \eta \nabla_\theta L\]
where \( \eta \) is the learning rate. By leveraging the low-rank structure, we reduce the number of trainable parameters and thus decrease the computational cost and memory overhead, while still enabling effective fine-tuning of the model.

Using this fine-tuning template, we fine-tune three separate models: using only the code generation dataset, only the code conversion dataset, and both datasets together. By fine-tuning on these datasets, the LLMs not only learns various code mappings between the source libraries and target libraries but also acquires knowledge of the target libraries. 

The task-specialized fine-tuned LLMs are then employed for seed code generation, detailed in Fig.~\ref{img:conv_and_gen}. On one hand, FUTURE leverages the code conversion capability of the fine-tuned LLMs to convert historical bug codes \begin{math}His\end{math} into potential bug codes \begin{math}Pot\end{math} for the target libraries. In the \begin{math}Pot\end{math}, there is a rich and diverse collection of complex and bug-prone API sequences that closely mimic real-world usage scenarios. On the other hand, FUTURE utilizes the code generation capability to generate diverse code snippets \begin{math}Gen\end{math} that call APIs of the target libraries. Together, \begin{math}Pot\end{math} and \begin{math}Gen\end{math} form the seed codes. During the acquisition of \begin{math}Pot\end{math} and \begin{math}Gen\end{math}, we perform automated preprocessing same as Section~\ref{sec:Historical Bug Collection}. Even though \begin{math}His\end{math} has been processed earlier, the inherent uncertainty of LLMs still necessitate this preprocessing to ensure the usability of the seed codes.

This strategy effectively transcends the limitations posed by relying solely on historical bug codes from the source libraries. It allows FUTURE to utilize historical bug information from source libraries, which TitanFuzz~\cite{titanfuzz} cannot do. Additionally, it facilitates the generation of code snippets of target libraries, even when models incorporating knowledge of these libraries are unavailable—something TitanFuzz is unable to achieve.

\subsection{Test Oracle}\label{sec:Test Oracle}
After obtaining seed codes, we implement differential testing, focusing on two critical aspects:

\textbf{Crash Bugs.} During the execution of the seed codes, we monitor for crashes such as aborts, segmentation faults, runtime errors, and floating point exceptions. Bugs exposed by such crashes may further trigger security vulnerabilities.
 
\textbf{Inconsistencies.} We scrutinize the execution results across different computational backends (CPU/GPU) for inconsistencies, which often indicate underlying bugs. Additionally, since each fuzzing seed code corresponds to an implementation in the source libraries, following~\cite{titanfuzz}, we calculate the Euclidean distance~\cite{euclidean} to measure the discrepancies between results. We only consider discrepancies exceeding a predefined threshold \( T \) as ``potential bugs”. Only when these discrepancies are reported to and confirmed by developers do we classify them as triggered bugs caused by inconsistencies between the source libraries (Src libs) and target libraries (Tar libs).

After conducting the test oracle, we report and perform a statistical analysis of the detected bugs. This analysis allows us to summarize and extract bug-prone API inputs, which are then used to automatically test the APIs of the source libraries. 

In FUTURE, there are only two manually involved parts. The first is extracting bug-prone API inputs during the bug reporting process. The second involves manually inspecting the causes of failures in code pairs generation, which constitute a small portion. This inspection helps us identify documentation issues related to certain APIs. Apart from these two parts, FUTURE is fully automated.

\section{Evaluation}

\subsection{Implementation}
\noindent \textbf{DL Library Selection.} We select PyTorch (v2.0.0)~\cite{pytorch} and TensorFlow (v2.13.0)~\cite{tensorflow} as the source libraries for FUTURE due to their extensive application. FUTURE can be applied to any newly introduced or prospective DL library. To assess its effectiveness, we focus on three target libraries introduced in recent years: Apple MLX (v0.0.10), Huawei MindSpore (v2.2.13), and OneFlow (v0.9.1). These open-source libraries represent the cutting-edge in DL library development, making them ideal candidates for evaluating FUTURE's effectiveness.

\begin{table}[t]
\caption{GitHub issues and bug codes collected.}
\label{tab:issues}
\begin{tabular*}{\linewidth}{@{\extracolsep{\fill}}cccc}
\toprule
\textbf{Source library}           & \textbf{Issue Label} & \textbf{Issue Number} & \textbf{Bug Codes} \\ \cmidrule{1-4}
\multirow{5}{*}{\textbf{PyTorch}} & bug                  & 315                   & 178                \\
                                  & Nans and Infs        & 138                   & 43                 \\
                                  & edge cases           & 184                   & 84                 \\
                                  & error checking       & 275                   & 86                 \\
                                  & crash                & 388                   & 139                \\ \cmidrule{1-4}
\textbf{Tensorflow}               & bug                  & 9975                  & 4503               \\ \cmidrule{1-4}
\textbf{ToTal}                    & -                    & 11275                 & 5033               \\ 
\bottomrule
\end{tabular*}
\end{table}

\noindent \textbf{Historical Bug Collection.} ~\label{sec:1530} FUTURE targets all bug-related issues up to March 20, 2024, processing a total of 11,275 issues from PyTorch and TensorFlow on GitHub. From these, we extract 5,033 valid historical bug codes, as detailed in Table~\ref{tab:issues}. Due to constraints in computational resources and time, we use the most recent 1,000 TensorFlow bug codes and all available PyTorch bug codes, resulting in a total of 1,530 historical bug codes utilized in this study.

\noindent \textbf{Large Language Models.} FUTURE utilizes CodeLlama-13B~\cite{codellama} for generating code pairs and fine-tuning, we select CodeLlama due to its advanced performance in open models and large input context support. The foundation model and its weights are sourced from Hugging Face~\cite{huggingface}, providing a cost-effective solution to users as it is available for free. 

\noindent \textbf{Fine-tuning Datasets.} We generate code pairs for 128, 156, and 168 APIs for MLX, MindSpore, and OneFlow respectively. For each API, we generate 5 code pairs, which are then mutated 100 times, resulting in a fine-tuning dataset of over six hundred thousand entries.

\noindent \textbf{Fine-tuning Setups.} During fine-tuning, we quantize the model's parameters to 4 bits. For the LoRA configuration, we follow the official tutorial provided by PEFT~\cite{peft}. As for the training parameters, the learning rate \( \eta \) is set at 3e-4 and the \begin{math}max\_steps\end{math} is set to 400. We allocate one-tenth of the training datasets for validation, conducting validations every 20 steps.

\subsection{Experimental Setup}
\noindent \textbf{Baselines.} Since the target libraries of FUTURE are newly introduced, most existing fuzzing methods are not readily applicable to all of them. As FUTURE is an API-level fuzzer, we focus on state-of-the-art API-level fuzzers such as FreeFuzz~\cite{freefuzz}, DeepREL~\cite{deeprel}, NablaFuzz~\cite{nablafuzz}, TensorScope~\cite{tensorscope}, and TitanFuzz~\cite{titanfuzz}, while also considering model-level fuzzers like Muffin~\cite{muffin} for a comprehensive understanding. Among these, FreeFuzz and DeepREL require extensive data collection and code modifications, making them resource-intensive for our needs. Muffin, limited to backends compatible with Keras, is not directly applicable to our target libraries without significant manual adaptations. Therefore, we select TitanFuzz, NablaFuzz and TensorScope as our baselines and adapt them to test our target libraries. In addition, to achieve a more comprehensive comparison, we selected few-shot learning~\cite{fewshot} as the fourth baseline to verify the effectiveness and rationality of fine-tuning in FUTURE through comparison.

All baseline fuzzers are modified minimally to ensure compatibility with our target libraries. We make several main adaptions as below: (1) replacing TitanFuzz's deprecated model with gpt-3.5-turbo, (2) updating API lists to match our targeted APIs, (3) modifying codes to save generated snippets.

In few-shot, for a fair comparison, we choose few-shot + CodeLlama (w/o fine-tuning) and we adopt chain-of-thought (CoT) prompting. For different tasks, we provide specific task descriptions along with 10-shot examples from our conversion or generation dataset, respectively. Using this context, we perform code conversion and generation accordingly.

\noindent \textbf{Environment.} We use an Ubuntu 20.04 server equipped with an Intel Xeon Gold 6130 CPU and a V100-32GB GPU.

\begin{table}[t]
\captionsetup{font={small}}
\caption{\textbf{Summary of CVEs detected by FUTURE.} All detected CVEs are found in OneFlow and due to the lack of strict parameter checking in APIs, indicating that OneFlow urgently needs to optimize the parameter validation mechanisms to prevent crashes or inconsistencies that could be exploited by attackers. }
\small
\label{tab:cves}
\begin{tabular*}{\hsize}{@{}@{\extracolsep{\fill}}lll@{}}
\toprule
\textbf{CVE ID}    & \textbf{Symptom}     & \textbf{Vulnerable API}                 \\ \cmidrule{1-3}
CVE-2024-36730 & Crash             & zeros/ones/new\_ones/empty \\
CVE-2024-36732 & Crash             & tensordot                       \\
CVE-2024-36734 & Crash             & var                             \\
CVE-2024-36735 & Src libs/Tar libs & eye                             \\
CVE-2024-36736 & Src libs/Tar libs & permute                         \\
CVE-2024-36737 & Crash             & full                            \\
CVE-2024-36740 & Crash             & scatter/scatter\_add            \\
CVE-2024-36742 & Crash             & scatter\_nd                     \\
CVE-2024-36743 & Crash             & dot                             \\
CVE-2024-36745 & Crash             & index\_select                   \\ \bottomrule
\end{tabular*}
\end{table}

\begin{table}[t]
\caption{Summary of bug detection on target libraries.}
\label{tab:bugs}
\begin{tabular*}{\hsize}{@{}@{\extracolsep{\fill}}ccccc@{}}
\toprule
                   & \multirow{2}{*}{\textbf{Total}} & \multicolumn{2}{c}{\textbf{Confirmed}} & \multirow{2}{*}{\textbf{Won’t fix}} \\ \cmidrule{3-4}
\textbf{}          &                                 & \textbf{Unknown}    & \textbf{Known}   &                                     \\ \cmidrule{1-5}
\textbf{MLX}       & 35                              & 32                  & 0                & 3                                   \\
\textbf{MindSpore} & 30                              & 22                  & 6                & 2                                   \\
\textbf{OneFlow}   & 83                              & 83                  & 0                & 0                                   \\
\textbf{Total}     & 148                             & 137                 & 6                & 5                                   \\     \bottomrule
\end{tabular*}
\end{table}

\subsection{Metrics}\label{sec:metrics}
\noindent \textbf{Number of Bugs.} We report potential bugs detected by FUTURE to developers via GitHub. We count only those bugs labeled as ``bug" as detected by FUTURE.

\noindent\textbf{Success rate of bug reproduction.} We define \begin{math}Suc\end{math} as converted codes that successfully reproduce the behaviors (error messages or outputs in bug reports) observed in the original bug codes. The success rate quantifies the effectiveness of FUTURE in code conversion and is calculated as $\mathsf{Success\ Rate}=\frac{N_{Suc}} {N_{His}}$, where \begin{math}N\end{math} represents the number of codes.
% FUTURE converts historical bug codes from the source libraries into potential bug codes of the target libraries. We define \begin{math}Suc\end{math} as converted codes that successfully reproduce the behaviors (error messages or outputs) observed in the original bug codes. The success rate quantifies the effectiveness of FUTURE in code conversion and is calculated as $\mathsf{Success\ Rate}=\frac{N_{Suc}} {N_{His}}$, where \begin{math}N\end{math} represents the number of codes.
% and \begin{math}His\end{math} are detailed in ~\ref{sec:Historical Bug Collection}.

\noindent\textbf{Validity rate.} For code generation, the validity rate is calculated as the ratio of the number of unique valid codes \begin{math}Val\end{math} to the total number of generated codes \begin{math}All\end{math}. A code is valid if it executes without exceptions and invokes the target API at least once. The validity rate is given as $\mathsf{Validity\ Rate} = \frac{N_{Val} }{N_{All}}$.

\noindent\textbf{API Coverage.} We measure API coverage by calculating the proportion of the APIs utilized in \begin{math}Suc\end{math} and \begin{math}Val\end{math}, relative to the total number of targeted APIs. Specifically, APIs that appear in \begin{math}Suc\end{math} are denoted as \(SucAPI\), and APIs that appear in \begin{math}Val\end{math} are denoted as \(ValAPI\). The API coverage is given by:

\begin{normalsize}
\begin{equation}
\mathsf{API\ Coverage} = \frac{(N_{SucAPI} \cup N_{ValAPI}) \cap N_{TarAPI}}{N_{TarAPI} } \nonumber
\end{equation}
\end{normalsize}
where \begin{math}N_{TarAPI}\end{math} represents the total number of targeted APIs.

\subsection{Research Questions}
To assess the effectiveness of FUTURE, we conduct studies answering the following research questions:

\begin{itemize}[leftmargin=0.5cm]
\item \textbf{RQ1:} Can FUTURE detect real-world bugs in newly-introduced DL libraries?
\item \textbf{RQ2:} How do key components and different settings of FUTURE influence its effectiveness?
\item \textbf{RQ3:} What is the fuzzing performance of FUTURE compared to the state-of-the-art techniques?
\end{itemize}

\section{Result Analysis}
\subsection{RQ1: Bug Detection}
We first investigate the effectiveness of FUTURE in bug detection. FUTURE detects 148 bugs across Apple MLX, Huawei MindSpore, and OneFlow, with 142 confirmed as previously unknown. Among these bugs, 10 have been assigned CVE IDs as detailed in Table~\ref{tab:cves}. Our submissions on GitHub are recognized with five ``good first issue" labels by MLX developers. Additionally, we identify 7 bugs in PyTorch. Since PyTorch is not one of our target libraries, we do not provide a detailed analysis of these bugs here. Statistics of bugs on the target libraries are presented in Table~\ref{tab:bugs}.

\begin{figure}[t]
  \setlength{\abovecaptionskip}{0cm}  %段前
  \setlength{\belowcaptionskip}{-0.2cm} %段后
  \centering
  \includegraphics[width=\columnwidth]{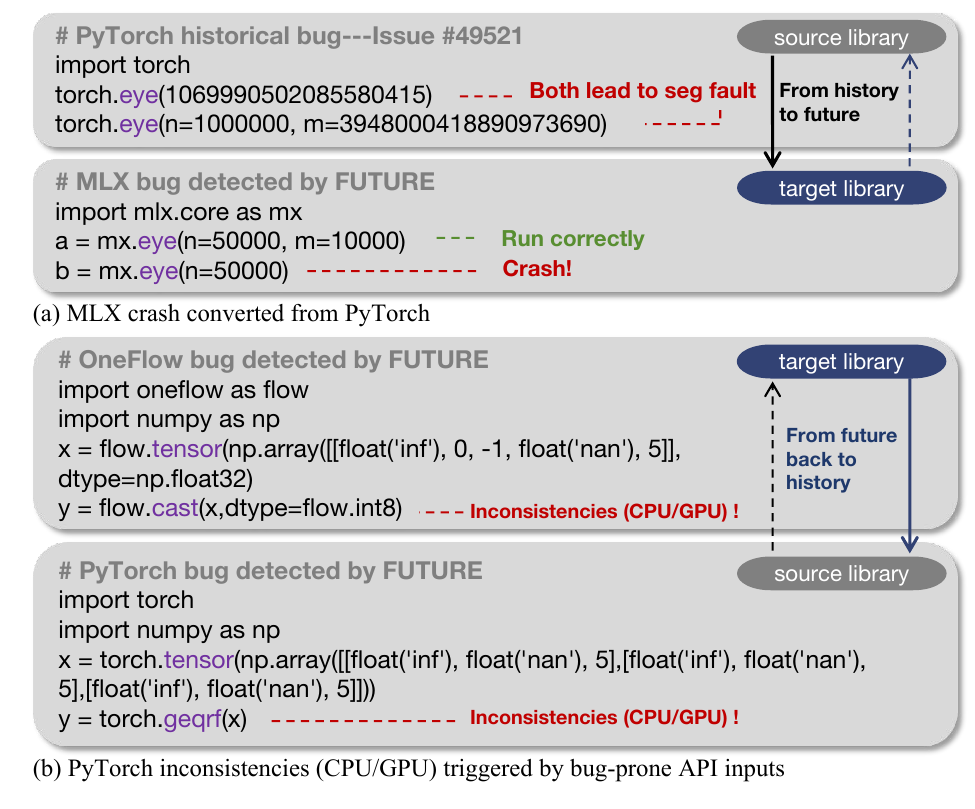}
  \caption{\textbf{Example bugs found by FUTURE.} We provide two bug examples to illustrate that FUTURE not only uses historical bug information from source libraries to unveil bugs in target libraries but also leverages bugs found in target libraries to identify bugs that still reside in source libraries.}
  \label{img:example}
\end{figure}

Fig.~\ref{img:example}(a) shows a historical bug code from PyTorch. In this bug, excessively large parameters passed to \begin{math}\mathsf{torch.eye}\end{math} trigger a segmentation fault. FUTURE converts this code into code snippets using target libraries. We find that on MLX, \begin{math}\mathsf{mlx.core.eye}\end{math} operates correctly when both parameters are set to large values. However, setting only one parameter to a large value triggers a crash. This issue is labeled as a bug and immediately fixed by MLX developers. This demonstrates that FUTURE can utilize historical bug information from existing libraries to unveil bugs in prospective libraries.

% \begin{table}[t]
% \setlength{\abovecaptionskip}{0cm}  %段前
% \setlength{\belowcaptionskip}{0.2cm} %段后
% % \captionsetup{font={small}}
% \caption{Documentation and Enhancement Finding.}
% \label{tab:documentation and enhancement}
% \begin{tabular*}{\hsize}{@{}@{\extracolsep{\fill}}ccc@{}}
% \toprule
% \multicolumn{1}{l}{} & \textbf{Enhancement} & \textbf{Documentation} \\ \hline
% \textbf{MLX}         & 13                   & 8                      \\
% \textbf{Mindspore}   & 7                    & 0                      \\
% \textbf{Oneflow}     & 0                    & 4                      \\
% \textbf{Total}       & 20                   & 12                     \\ \bottomrule
% \end{tabular*}
% \end{table}

\begin{table}
\caption{Causes of bug detected on target libraries.}
\label{tab:causes}
\begin{tabular*}{\hsize}{@{}@{\extracolsep{\fill}}ccccccc@{}}
\toprule
\multicolumn{1}{l}{} & \textbf{Total} & \textbf{EC} & \textbf{NI} & \textbf{LD} & \textbf{DBI} & \textbf{MPC} \\ \cmidrule{1-7}
\textbf{MLX}         & 35             & 4                   & 17                     & 3                         & 4                                          & 7                                   \\
\textbf{MindSpore}   & 30             & 0                   & 25                     & 0                         & 1                                          & 4                                   \\
\textbf{OneFlow}     & 83             & 0                   & 51                     & 3                         & 1                                          & 28                                  \\
\textbf{Total}       & 148            & 4                   & 
\textbf{93}\                    & 6                         & 6                                          & \textbf{39}\                                 \\ \bottomrule
\end{tabular*}
\end{table}

Fig.~\ref{img:example}(b) illustrates a bug detected by FUTURE through its capability to generate random codes for target libraries. We find that arrays containing NaNs and Infs can trigger many bugs in target libraries. By extracting bug-prone API inputs and testing them on the APIs of source libraries, we identify that \begin{math}\mathsf{torch.geqrf}\end{math} exhibits inconsistencies on CPU and GPU. This issue persists in the latest version of PyTorch and has been confirmed as a bug. This demonstrates that FUTURE uses insights from prospective libraries to enhance security in existing ones, creating a cycle from history to future and back.

Beyond bugs, 20 issues detected by FUTURE, though not classified as bugs, are recognized by developers as valuable enhancement issues. Additionally, FUTURE identifies 12 documentation problems, we report them and they have all been addressed in the latest versions of the official documentation. 
% Detailed statistics on the documentation and enhancement issues are presented in Table \ref{tab:documentation and enhancement}.

\noindent\textbf{Bug Cause Analysis.} Based on labels of historical bugs and the actual circumstances of bugs detected by FUTURE, we classify the causes of the detected bugs into five categories:
\begin{itemize}[leftmargin=0.5cm]
\item \textbf{Nans and Infs (NI):} Bugs in this category involve the presence of NaNs or Infs in two specific contexts: as variables in an input array or as API parameters.
\item \textbf{Missing Parameter Checking (MPC):} These bugs result from insufficient or flawed parameter validation within APIs, allowing parameters that do not meet constraints to pass checks, leading to incorrect results or crashes.
\item \textbf{Edge Cases (EC):} These bugs involve bugs in APIs when handling boundary values for certain data types.
\item \textbf{Different Backend Implementations (DBI):} Bugs in this category arise from discrepancies in the implementation of DL libraries across different backends (CPU/GPU), affecting the outputs under identical inputs and parameters.
\item \textbf{Logic Deficiency (LD):} Some APIs suffer from logic deficiencies, preventing them from functioning as intended.
\end{itemize}

\begin{figure}[t]
  \centering
  \includegraphics[width=\columnwidth]{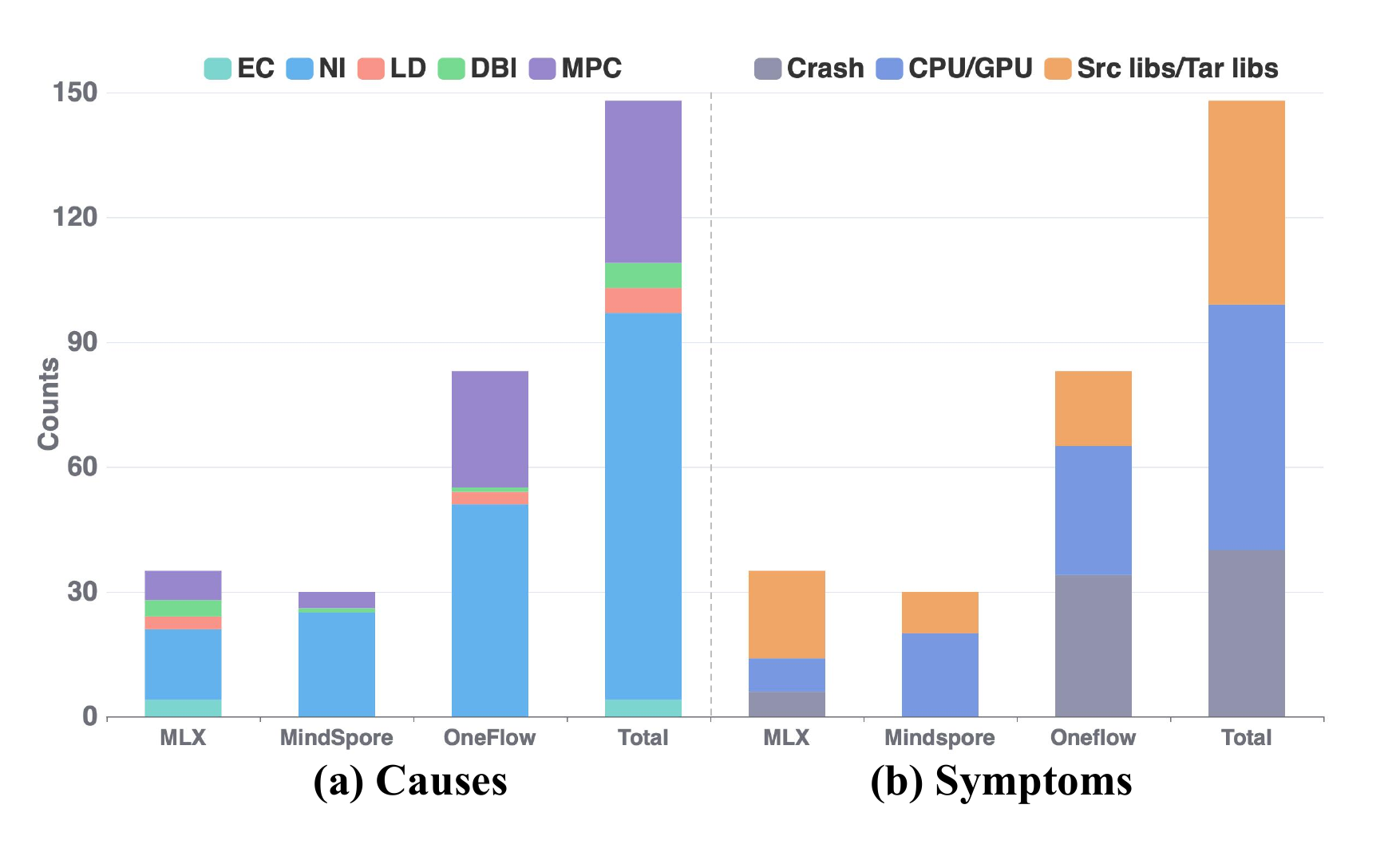}
  \caption{Statistical distribution of causes and symptoms of bugs detected in target libraries using FUTURE.}
  \label{img:causes}
\end{figure}

\begin{table}
\setlength{\tabcolsep}{4.7pt}
\caption{Symptoms of bug detected on target libraries.}
\label{tab:symptoms}
\begin{tabular*}{\hsize}{@{}@{\extracolsep{\fill}}ccccc@{}}
\toprule
\multicolumn{1}{l}{} & \textbf{Total} & \textbf{Crash} & \textbf{CPU/GPU} & \textbf{Src libs/Tar libs} \\ \cmidrule{1-5}
\textbf{MLX}         & 35             & 6              & 8                                 & 21                                          \\
\textbf{MindSpore}   & 30             & 0              & 20                                & 10                                          \\
\textbf{OneFlow}     & 83             & 34             & 31                                & 18                                          \\
\textbf{Total}       & 148            & 40             & 59                                & 49                                          \\ \bottomrule
\end{tabular*}
\end{table}

We categorize the bugs detected by FUTURE according to these causes and compile statistics. Table~\ref{tab:causes} details the number of bugs per category, and Fig.~\ref{img:causes}(a) illustrates the distribution of each cause within the bugs detected in each target library. \emph{For bugs involving inputs or parameters with NaNs and Infs, and where outputs vary across different backends, we categorize the cause as ``NI"}. Only when a bug does not involve NaNs and Infs and there are discrepancies in outputs across different backends do we classify it as ``DBI". Similarly, when NaNs and Infs used as parameters trigger a bug, we prioritize classifying it under ``NI" rather than ``MPC".

Among all bugs, ``NI" is the most common cause, accounting for 62.83\% (93 out of 148). ``MPC" is the second most common cause, accounting for 26.3\% (39 out of 148), suggesting stricter parameter checking is needed during the development of DL libraries.

\noindent\textbf{Bug Symptom Analysis.} We also categorize the symptoms of the bugs based on the design of test oracle. Table~\ref{tab:symptoms} and Fig.~\ref{img:causes}(b) illustrate the distribution of bugs according to symptom categories, divided into three types:

\begin{itemize}[leftmargin=0.5cm]
\item \textbf{Crash:} This category includes scenarios where seed code execution results in aborts, segmentation faults, runtime errors, floating point exceptions, and excessively long execution times without producing results or errors.
\item \textbf{Inconsistencies (CPU/GPU):} These bugs are characterized by different outcomes for the same API executed on different backends, despite identical inputs and parameters.
\item \textbf{Inconsistencies (Src libs/Tar libs):} These bugs are identified when the outcomes of the target library seed codes are consistent across different backends but differ from those of corresponding implementations in the source libraries, given the same inputs and parameters.
\end{itemize}

Our findings reveal that OneFlow has a higher incidence of bugs manifesting as crashes, accounting for 40.96\% (34 out of 83) of the total bugs detected in OneFlow by FUTURE. This is likely due to inadequate handling of invalid inputs or parameters in OneFlow's source code, often leading to program terminations. Additionally, FUTURE finds no crash-related bugs in MindSpore. We attribute this to MindSpore's implementation of stricter parameter checks and more robust error handling mechanisms.

\begin{tcolorbox}[size=title,boxsep=1pt,left=1pt,right=1pt,top=0pt,bottom=0pt]
\textbf{Answer to RQ1:} FUTURE detects 148 bugs across MLX, MindSpore, and OneFlow, with 142 confirmed as previously unknown. The analysis on bug causes and symptoms provides valuable insights for further research in DL libraries.
\end{tcolorbox}

\subsection{RQ2: Impact of Key Components and Different Settings}

\begin{table*}[htbp]
% \small
\setlength{\tabcolsep}{4.3pt}
\captionsetup{font={small}}
\caption{\textbf{Evaluation of historical bug collection.} In the brackets following the percentages, the numerator and denominator for the success rate are represented by \begin{math}N_{Suc}\end{math} and \begin{math}N_{His}\end{math} respectively; for API coverage, they are covered API and all targeted API. The detailed definitions of these metrics can be found in Section~\ref{sec:metrics}. The column \textbf{Bug detected} shows the number of bugs detected by FUTURE using historical bug codes with specific labels.}
\label{tab:history}

% for the validity rate, they are \begin{math}N_{Val}\end{math} and \begin{math}N_{All}\end{math}

\begin{tabular}{ccccccccc}
\toprule
\multicolumn{2}{c}{\textbf{Historical Bug Codes on Source Libraries}}            & \multicolumn{7}{c}{\textbf{Converted Codes on Target Libraries}}                                                                                               \\ \cmidrule(r){1-2} \cmidrule(r){3-9}
\multirow{2}{*}{\textbf{Source library}} & \multirow{2}{*}{\textbf{Issue Label}} & \multirow{2}{*}{\textbf{Success Rate}} & \textbf{}      & \textbf{API Coverage} & \multicolumn{1}{l}{} & \multicolumn{3}{c}{\textbf{Bug detected}}            \\ \cmidrule{4-9} 
                                         &                                       &                                        & \textbf{MLX}   & \textbf{MindSpore}    & \textbf{OneFlow}     & \textbf{MLX} & \textbf{MindSpore} & \textbf{OneFlow} \\ \cmidrule(r){1-2} \cmidrule(r){3-9}
                      & bug                                   & 11.8\% (21/178)                         & 17.2\% (22/128) & 9.0\% (14/156)         & 11.3\% (19/168)       & 2            & 1                  & 5                \\
\textbf{}                                & Nans and Infs                         & \textbf{20.9\% (9/43)}                           & \textbf{25.0\% (32/128)} & \textbf{26.3\% (41/156)}        & \textbf{28.0\% (47/168)}       & \textbf{6}            & \textbf{4}                  & \textbf{11}               \\       \textbf{PyTorch}                     & edge cases                            & 13.1\% (11/84)                          & 21.1\% (27/128) & 16.0\% (25/156)        & 7.7\% (13/168)        & 1            & 0                  & 1                \\
\textbf{}                                & error checking                        & 15.1\% (13/86)                          & 8.6\% (11/128)  & 12.2\% (19/156)        & 14.9\% (25/168)       & 0            & 1                  & 1                \\
\textbf{}                                & crash                                 & 11.5\% (16/139)                         & 13.3\% (17/128) & 21.8\% (34/156)        & 18.5\% (31/168)       & 2            & 0                  & 2                \\ \cmidrule(r){1-2} \cmidrule(r){3-9}
\textbf{TensorFlow}                      & bug                                   & 8.6\% (86/1000)                         & 34.4\% (44/128) & 35.9\% (56/156)        & 45.2\% (76/168)       & 2            & 1                  & 5                \\ \cmidrule(r){1-2} \cmidrule(r){3-9}
\textbf{Total}                           & -                 & 10.2\% (156/1530)                       & 57.0\% (73/128) & 65.4\% (102/156)       & 68.5\% (115/168)      & 13           & 7                  & 25               \\ \bottomrule
\end{tabular}
\end{table*}

\begin{table*}[ht]
\caption{Evaluation of different methods for constructing the fine-tuning dataset.}
\label{tab:code pairs}
\begin{threeparttable}
\begin{tabular*}{\hsize}{@{}@{\extracolsep{\fill}}ccccccc@{}}
\toprule
\multicolumn{1}{l}{} & \multirow{2}{*}{\textbf{Success Rate}} & \multirow{2}{*}{\textbf{Validity Rate}} & \multicolumn{4}{c}{\textbf{API Coverage}}                                 \\ \cmidrule{4-7} 
\textbf{}            &                                        &                                         & \textbf{MLX}    & \textbf{MindSpore} & \textbf{OneFlow} & \textbf{Total}  \\ \cmidrule{1-7}
\textbf{FUTURE}      & \textbf{10.2\% (156/1530)}                       & \textbf{96.5\% (2894/3000)}                       & \textbf{98.4\% (126/128)} & \textbf{98.7\% (154/156)}    & \textbf{97.0\% (163/168)}  & \textbf{98.0\% (443/452)} \\
\textbf{FUTURE-doc}  & 3.4\% (52/1530)                         & 45.8\% (1375/3000)                       & 35.2\% (45/128)  & 37.2\% (58/156)     & 41.1\% (69/168)   & 38.1\% (172/452) \\
\textbf{FUTURE-ex}   & 4.4\% (68/1530)                         & 49.6\% (1487/3000)                       & 40.6\% (52/128)  & 52.0\% (81/156)     & 38.1\% (64/168)   & 43.6\% (197/452) \\
\textbf{FUTURE-no}   & 1.4\% (21/1530)                         & 36.1\% (1083/3000)                       & 28.9\% (37/128)  & 28.2\% (44/156)     & 19.6\% (33/168)   & 25.2\% (114/452)\\ \bottomrule
\end{tabular*}
\begin{tablenotes}
    \footnotesize
    \item[1] \textbf{FUTURE} represents fine-tuning with datasets constructed by generating and mutating code pairs. \textbf{FUTURE-doc} relies solely on API documentation and \textbf{FUTURE-ex} directly uses code examples to fine-tune CodeLlama. \textbf{FUTURE-no} represents the version without fine-tuning.
\end{tablenotes}
\end{threeparttable}
\end{table*}

To study how each component of FUTURE contributes to the overall effectiveness, we conduct experiments based on the four key components: (1) historical bug collection, (2) dataset construction, (3) fine-tuning, and (4) seed code generation.

\subsubsection{Historical Bug Collection} We perform code conversion on 1,530 historical bug codes mentioned in Section~\ref{sec:1530}, with the detailed results shown in Table~\ref{tab:history}.

\begin{figure}
  \centering
  \includegraphics[width=\columnwidth]{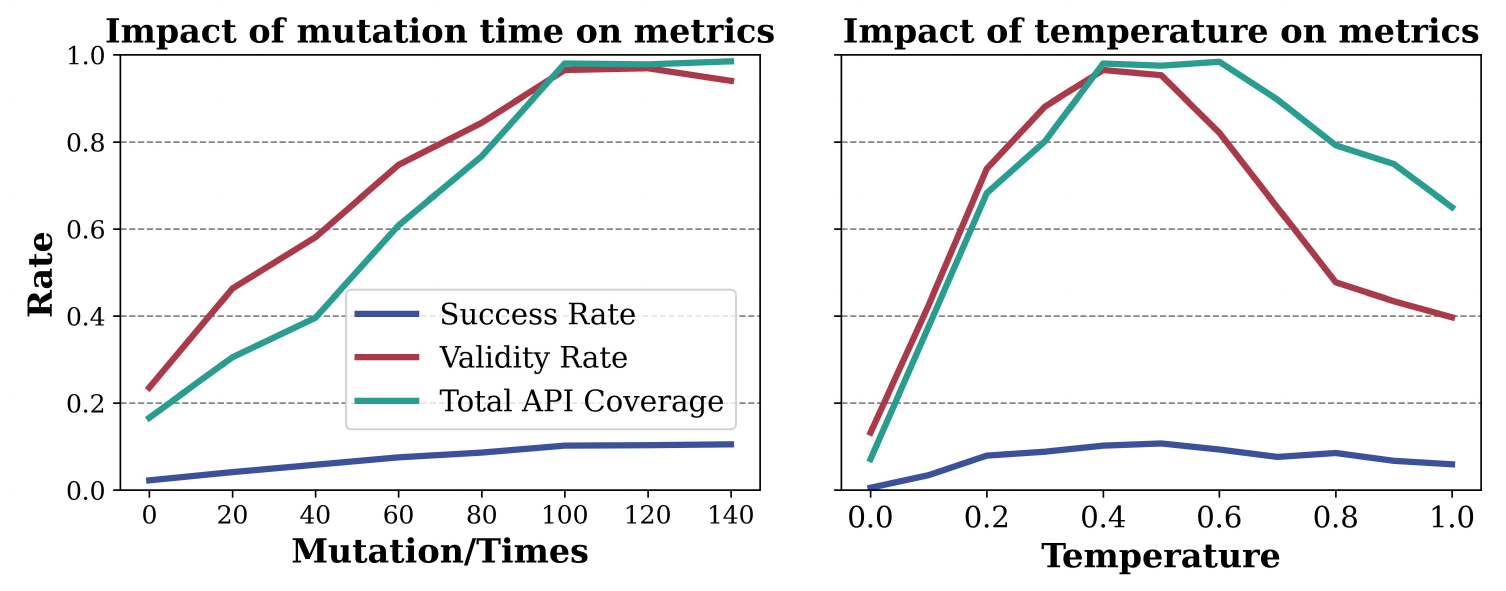}
  \caption{Impact of various mutation times and temperature on success rate, validity rate, and total API coverage.}
  \label{img:mutate}
\end{figure}

First, we observe that issues labeled ``Nans and Infs" in PyTorch provide the most extensive API coverage and are instrumental in bug detection. This observation confirms our prompt design strategies as discussed in Section~\ref{sec:GPT Prompt Construction}. Second, we note that the success rate of bug reproduction for TensorFlow's historical bug codes is notably low at 8.6\%. This is likely due to the vast number of TensorFlow APIs, many of which are undeveloped in the target libraries after conversion. Despite this, the historical bug codes of TensorFlow contribute significantly to overall API coverage and assist in detecting 8 bugs within three test libraries.

In summary, our experiments demonstrate that utilizing historical bug information from existing DL libraries contributes to detecting bugs in prospective DL libraries. As the collection of bugs in existing DL libraries continues to accumulate, this contribution is expected to become increasingly effective. Among the 45 bugs detected by FUTURE using historical bug information, 33 involved multi-API sequences, proving that FUTURE effectively leverages bug-prone API sequences rather than randomly generating API sequences with LLMs, as seen in TitanFuzz.

\subsubsection{Dataset Construction}\label{sec:last} Next, we examine how constructing and leveraging datasets for fine-tuning with different strategies influences the effectiveness of FUTURE.

\noindent \textbf{How to Construct Datasets.} We compare several variants of constructing datasets with different strategies. Table~\ref{tab:code pairs} shows the performance metrics of these strategies.

We observe that constructing the fine-tuning dataset by generating and mutating code pairs significantly enhances the success rate, validity rate and API coverage compared to other strategies. Notably, the metrics of \textbf{FUTURE-ex} outperform \textbf{FUTURE-doc}, indirectly confirming that using code snippets as datasets more effectively simulates real-world code conversion and generation scenarios.

We evaluate the impact of varying mutation times on the performance metrics, shown in Fig.~\ref{img:mutate}(a). We find that mutating code pairs significantly improves all metrics. However, when increasing mutations beyond 100, the benefits do not justify the additional time and resources needed. This finding guides us to cap mutations at 100 per code pair.

\noindent \textbf{How to Leverage Datasets.} To equip LLMs with the capabilities for code conversion and code generation, we construct two specialized datasets. We first investigate the impact of utilizing different combinations of these datasets on the effectiveness of FUTURE's code generation capability. For each target library, we generate 1,000 random code snippets for evaluation.

\begin{table}
\captionsetup{font={small}}
\caption{The impact of different dataset combinations used for fine-tuning on FUTURE's code generation effectiveness.}
\label{tab:generation}
\begin{threeparttable}
\begin{tabular*}{\hsize}{@{}@{\extracolsep{\fill}}ccc@{}}
\toprule
\multicolumn{1}{l}{} & \textbf{Validity Rate} & \textbf{Total API Coverage} \\ \midrule
\textbf{FUTURE}      & \textbf{96.5\% (2894/3000)}      & \textbf{92.3\% (417/452)}             \\
\textbf{FUTURE-gen}  & 81.1\% (2432/3000)      & 83.8\% (379/452)             \\
\textbf{FUTURE-conv} & 52.5\% (1576/3000)      & 56.2\% (254/452)             \\
\textbf{FUTURE-no}   & 36.1\% (1083/3000)      & 22.8\% (103/452)             \\ \bottomrule
\end{tabular*}
\begin{tablenotes}
    \footnotesize
    \item [1] \textbf{FUTURE} uses both datasets, \textbf{FUTURE-gen} performs fine-tuning with only code generation dataset, \textbf{FUTURE-conv} employs only code conversion dataset, and \textbf{FUTURE-no} indicates no fine-tuning at all.
\end{tablenotes}
\end{threeparttable}
\end{table}

\begin{table}
\captionsetup{font={small}}
\caption{\textbf{The impact of different dataset combinations on FUTURE's code conversion effectiveness.} All the variants of FUTURE remain consistent with the configurations in Table~\ref{tab:generation}.}
\label{tab:conversion}
\begin{tabular*}{\hsize}{@{}@{\extracolsep{\fill}}ccc@{}}
\toprule
\multicolumn{1}{l}{} & \textbf{Success Rate} & \textbf{Total API Coverage} \\ \midrule
\textbf{FUTURE}      & \textbf{10.2\% (156/1530)}      & \textbf{94.5\% (290/452)}             \\
\textbf{FUTURE-gen}  & 5.0\% (77/1530)        & 32.7\% (148/452)             \\
\textbf{FUTURE-conv} & 8.6\% (132/1530)       & 52.0\% (235/452)             \\
\textbf{FUTURE-no}   & 1.4\% (21/1530)        & 13.5\% (61/452)              \\ \bottomrule
\end{tabular*}
\end{table}

Table~\ref{tab:generation} presents the impact of different dataset combinations on FUTURE’s code generation apability. \textbf{FUTURE-gen} achieves significantly higher validity rate and API coverage compared to \textbf{FUTURE-conv} and \textbf{FUTURE-no}, indicating that the code generation dataset substantially enhances FUTURE's capability to generate relevant and effective code snippets.

Next, we conduct experiments to evaluate the impact of these different dataset combinations on FUTURE's code conversion capability using the same four versions of FUTURE described above. We perform library conversion on the 1,530 historical bug codes mentioned in Section~\ref{sec:1530}. The experimental results are detailed in Table~\ref{tab:conversion}. These results demonstrate that the code conversion dataset can enhance FUTURE's success rate in converting code between libraries. Using both of the designed datasets maximizes their benefits, affirming their complementary nature.

\subsubsection{Fine-tuning} We further assess the effectiveness of fine-tuning CodeLlama by comparing metrics with and without fine-tuning. Table~\ref{tab:code pairs} presents the performance enhancements achieved through fine-tuning. The results show that \textbf{FUTURE} significantly outperforms \textbf{FUTURE-no}, with improvements exceeding threefold across all metrics, demonstrating the substantial benefits of further fine-tuning on general-purpose fine-tuned LLMs. Even with a model like CodeLlama-13B, which has limited performance, FUTURE still achieved notable results, highlighting the framework's effectiveness regardless of the underlying model's capabilities.

\subsubsection{Seed Code Generation}\label{sec:difsetting}
The main goal of seed code generation is to implement code conversion and generation accurately. Therefore, we examine the impact of the \begin{math}temperature\end{math} hyperparameter on FUTURE's various performance. Fig.~\ref{img:mutate}(b) shows the metrics of FUTURE at varying \begin{math}temperature\end{math} settings. We observe that with the \begin{math}temperature\end{math} set to the default value of 0.4, FUTURE demonstrates superior capabilities in code conversion and code generation, effectively maximizing coverage across targeted APIs. 

\begin{tcolorbox}[size=title,boxsep=1pt,left=1pt,right=1pt,top=0pt,bottom=0pt]
\textbf{Answer to RQ2:} The effectiveness of FUTURE is enhanced by its key components, which significantly boost success and validity rates, as well as API coverage and bug detection. 
\end{tcolorbox}

\begin{table}[t]
\setlength{\abovecaptionskip}{0cm}  %段前
\setlength{\belowcaptionskip}{-0.2cm} %段后
\setlength{\tabcolsep}{1.4pt}
\scriptsize
\caption{Comparison on API coverage.}
\label{tab:covcomparison}
\begin{tabular*}{\hsize}{@{}@{\extracolsep{\fill}}ccccc@{}}
\toprule
                   & \textbf{MLX}    & \textbf{MindSpore} & \textbf{OneFlow} & \textbf{Total}  \\ \midrule
\textbf{FUTURE}    & \textbf{98.4\% (126/128)} & \textbf{98.7\% (154/156)}    & \textbf{97.0\% (163/168)}  & \textbf{98.0\% (443/452)} \\
\textbf{TitanFuzz} & 29.7\% (38/128)  & 89.1\% (139/156)    & 82.9\% (126/168)  & 67.0\% (303/452) \\
\textbf{NablaFuzz} & -               & -                  & 78.0\% (131/168)  & 28.9\% (131/452) \\
\textbf{TensorScope} & -             & 30.1\% (47/156)     & -                & 10.4\% (47/452) \\
\textbf{Few-shot} & 59.4\% (76/128)             & 77.56\% (121/156)     & 81.5\% (137/168)                & 73.9\% (334/452) \\
\bottomrule
\end{tabular*}
\end{table}

\begin{table}
\captionsetup{font={small}}
\caption{\textbf{Comparison on code generation.} In the table, for individual target library, the denominators for the percentages are all 1,000. They are omitted due to space limitations.}
\label{tab:gencomparison}
\begin{tabular*}{\hsize}{@{}@{\extracolsep{\fill}}ccccc@{}}
\toprule
\multicolumn{1}{l}{} & \textbf{MLX} & \textbf{MindSpore} & \textbf{OneFlow} & \textbf{Total} \\ \midrule
\textbf{FUTURE}      & \textbf{97.5\%}      & \textbf{95.2\%}            & \textbf{96.7\%}          & \textbf{96.5\%}        \\
\textbf{FUTURE-gen}  & 83.4\%      & 79.8\%            & 80.0\%          & 81.1\%        \\
\textbf{FUTURE-conv} & 54.2\%      & 44.3\%            & 59.1\%          & 52.5\%        \\
\textbf{FUTURE-no}   & 14.2\%      & 45.3\%            & 48.8\%          & 36.1\%        \\
\textbf{TitanFuzz}   & 15.7\%      & 41.5\%            & 52.7\%          & 36.6\%        \\
\textbf{Few-shot} & 42.7\%      & 70.5\%            & 63.9\%          & 59.0\%        \\
\bottomrule
\end{tabular*}
\end{table}

\subsection{RQ3: Comparison with Other Work}
In this section, we compare FUTURE with four baselines: TitanFuzz, NablaFuzz, TensorScope and few-shot learning. 
% We use TitanFuzz to generate 10 code snippets for each targeted API, maintaining all other settings at their default configurations.

We did not find any confirmed bugs using our baselines. Table~\ref{tab:covcomparison} presents the API coverage of all evaluated baselines on our target libraries. NablaFuzz, which does not support MLX and MindSpore and relies on its proprietary database inaccessible to us, has its performance recorded only for OneFlow. Similarly, TensorScope's design, which involves constraints among APIs of multiple libraries, makes migration to MLX and OneFlow challenging with minimal effort.

We observe that FUTURE achieves significantly higher API coverage across all target libraries, with an increase of 46.2\% compared to TitanFuzz. TitanFuzz performs relatively well on MindSpore and OneFlow but poorly on MLX, likely due to the training data of gpt-3.5-turbo not including information of MLX. This highlights FUTURE's suitability for prospective libraries, where existing fuzzers like TitanFuzz may be ineffective. Few-shot learning addresses this issue to some extent, but its performance still lags significantly behind fine-tuning. FUTURE achieves a 24.4\% improvement on OneFlow compared to NablaFuzz and outperforms TensorScope on newer versions of MindSpore.

TitanFuzz first realizes a fully-automated framework to perform generation-based fuzzing directly leveraging LLMs. We conduct experiments comparing the code generation capability of different versions of FUTURE with TitanFuzz and few-shot learning. For each target library, we generate 1,000 code snippets. Table~\ref{tab:gencomparison} presents the validity rate of the snippets.

FUTURE significantly outperforms TitanFuzz in terms of validity rates across all target libraries. Specifically, the total validity rate of FUTURE reaches 263.7\% of TitanFuzz. For MLX, the validity rate is more than six times higher than TitanFuzz. \textit{TitanFuzz's performance is only comparable to} \textbf{FUTURE-no}. These findings indicate that, with the  FUTURE's code generation capability for newly introduced and prospective DL libraries significantly surpasses that of the state-of-the-art generation-based DL library fuzzing method.

Few-shot learning shows an improved validity rate across all libraries compared to TitanFuzz, but it still falls short of FUTURE. Through manual analysis of some of the code snippets generated by few-shot learning, we find that the quality of the generated code snippets deteriorate over time, with instances where the generated code even failed to adhere to the instructions in the initial prompt. In contrast, after fine-tuning, each generated code snippet is a new invocation of the fine-tuned model, which effectively avoids the issues encountered in few-shot learning.

\begin{tcolorbox}[size=title,boxsep=1pt,left=1pt,right=1pt,top=0pt,bottom=0pt]
\textbf{Answer to RQ3:} FUTURE consistently outperforms baselines in multiple metrics, showcasing its superior efficacy and adaptability for newly introduced and prospective libraries.
\end{tcolorbox}

\section{Threats to Validity}
\noindent\textbf{Internal:} In our experiments, we employ CodeLlama to generate code pairs based on API documentation and code examples. Given the inherent uncertainties and variability in the performance of LLMs, not all generated code pairs may meet the desired quality standards. We mitigate this uncertainty through designing rigorous preprocessing and validity checks on the code pairs. By implementing these checks, we aim to reduce the impact of low-quality code pairs and maintain the reliability of our experimental results.

\noindent\textbf{External:} Applying FUTURE to DL libraries not initially targeted by the framework may compromise its effectiveness, particularly if users are unfamiliar with the fine-tuning process of LLMs. To address this threat, we provide a comprehensive fine-tuning template that guides users through the adaptation process. By offering this resource, we aim to empower users to effectively customize FUTURE to their specific needs with minimal effort, thereby enhancing its applicability and robustness across various DL libraries.

\section{Conclusion}
In this work, we propose FUTURE, the first universal DL library fuzzing framework designed for both newly introduced and prospective DL libraries. More specifically, FUTURE collects historical bug codes from existing libraries, fine-tunes LLMs with limited available information. With the historical bug codes and fine-tuned LLMs, we generate seed codes and perform differential testing, enhancing security in both new and existing libraries. Our evaluation on three newly introduced libraries shows that FUTURE significantly outperforms existing fuzzers in multiple dimensions. Notably, FUTURE has detected 148 bugs across 452 targeted APIs, including 142 previously unknown bugs. Among these bugs, 10 have been assigned CVE IDs. Our submissions on GitHub are recognized with five ``good first issue" labels by MLX developers. Additionally, FUTURE detects 7 bugs in PyTorch, demonstrating the framework’s ability to utilize historical bug information to secure new libraries and enhance existing ones in reverse. In subsequent research, we aim to expand the range of FUTURE's source libraries, incorporating a broader spectrum of historical bug information from various DL libraries. Additionally, we will implement more automated components to complete the cycle from future back to history.

\section*{Acknowledgment}
We sincerely appreciate all the anonymous reviewers for their insightful and constructive comments to improve this work. This paper is supported by the the Strategic Priority Research Program of the Chinese Academy of Sciences under No. XDA0320401, the National Natural Science Foundation of China under No. 62202457, and the Open Source Community Software Bill of Materials (SBOM) Platform under No. E3GX310201. This paper is supported by YuanTu Large Research Infrastructure.

\bibliographystyle{IEEEtran}
\bibliography{software}

\end{document}